\documentclass[aps,prb,twocolumn,showpacs,superscriptaddress,preprintnumbers]{revtex4}
\usepackage{epsfig,color}
\usepackage{bm}
\usepackage{graphicx}

\begin{document}

\title{Lifshitz quantum phase transitions and Fermi surface transformation with hole doping in high-$T_c $ superconductors}

\author{S.G.~Ovchinnikov}
 \affiliation{L.V. Kirensky Institute of Physics, Siberian Branch of Russian Academy of Sciences, 660036 Krasnoyarsk, Russia}
 \affiliation{Siberian Federal University, Krasnoyarsk, 660041, Russia}

\author{M.M.~Korshunov}
 \affiliation{L.V. Kirensky Institute of Physics, Siberian Branch of Russian Academy of Sciences, 660036 Krasnoyarsk, Russia}
 \affiliation{Max-Planck-Institut f\"{u}r Physik komplexer Systeme, D-01187 Dresden, Germany}
 \affiliation{Department of Physics, University of Florida, Gainesville, Florida 32611, USA}

\author{E.I.~Shneyder}
 \email{shneyder@iph.krasn.ru}
 \affiliation{L.V. Kirensky Institute of Physics, Siberian Branch of Russian Academy of Sciences, 660036 Krasnoyarsk, Russia}
 \affiliation{Reshetnev Siberian State Aerospace University, Krasnoyarsk 660014, Russia}

\date{\today}

\begin{abstract}
We study the doping evolution of the electronic structure in the normal phase
of high-$T_c$ cuprates. Electronic structure and Fermi surface of cuprates with
single CuO$_2$ layer in the unit cell like La$_{2-x}$Sr$_x$CuO$_4$ have been
calculated by the LDA+GTB method in the regime of strong electron correlations
(SEC) and compared to ARPES and quantum oscillations data. We have found two
critical concentrations, $x_{c1}$ and $x_{c2}$, where the Fermi surface
topology changes. Following I.M. Lifshitz ideas of the quantum phase
transitions (QPT) of the $2.5$-order we discuss the concentration dependence of
the low temperature thermodynamics. The behavior of the electronic specific
heat $\delta(C/T) \sim \left( x - x_c \right)^{1/2}$ is similar to the Loram
and Cooper experimental data in the vicinity of $x_{c1} \approx 0.15$.
\end {abstract}


\maketitle

\section{Introduction}

Nowadays high-$T_c$ cuprates is the second most studied class of condensed
matter after semiconductors. Both the nature of the superconductivity and the
abnormal pseudogap feature of the normal phase are not clear
yet~\cite{r1,r2,r3,r4,r5,r6,r7,r8}. A lot of experimental data on the
electronic structure have been obtained by ARPES that reveals the doping
evolution of the Fermi surface (FS) from small arcs near $(\pi/2,\pi/2)$ at
small doping to the large FS around $(\pi,\pi)$ at large doping~\cite{r9}.
Quantum oscillations measurements in strong magnetic fields on the single
crystals YBa$_2$Cu$_3$O$_{6.5}$~\cite{r10} and YBa$_2$Cu$_4$O$_8$~\cite{r11}
have proved the existence of small hole pockets in the underdoped (UD) cuprates
that looks as a contradiction to the ARPES arcs. This contradiction has been
explained by the interaction between holes and spin fluctuations in the
pseudogap state with the existing short-range antiferromagnetic (AFM)
order~\cite{r12,r13,r14,r15}. It occurs that the part of the hole pocket
related to the shadow band has smaller quasiparticle (QP) lifetime due to the
QP scattering on spin fluctuations. Recently VUV laser ARPES~\cite{r70} has
found a closed FS pocket in the UD La-Bi2201 with the small intensity at the
shadow band part. The strong interaction of the electrons with the spin
fluctuations is a general property of SEC systems and takes place not only in
cuprates but also \textit{e.g.} in manganates~\cite{r16}.

The conventional LDA (local density approximation) approach to the electronic
structure in the regime of SEC fails. Various realistic multiband models of
CuO$_2$ layer in cuprates in the low energy region result in the effective
Hubbard and $t-J$ models~\cite{r17,r18,r19,r20,r21}. In the hybrid LDA+GTB
scheme~\cite{r22} that combines the LDA calculations of the multiband $p-d$
model parameters and the generalized tight-binding (GTB) treatment of SEC the
low energy effective $t-t'-t''-J^*$ model has been obtained from microscopic
approach with all parameters being calculated {\it ab initio}.

Small hole pockets in the UD case with area $\sim x$ appear in a theory
considering the hole dynamics in the AFM spin background and have been obtained
by the exact diagonalization~\cite{r23} and the Quantum Monte Carlo studies of
the finite clusters~\cite{r24,r25} as well as by various variational and
perturbation calculations for the infinite-dimension
lattice~\cite{r26,r27,r28,r29,r30}. Once the long-range AFM order disappears
with doping the electronic structure calculations in the paramagnetic phase
results in the dispersion of the valence band with the top at $(\pi,\pi)$ and
the large FS~\cite{r31}. Still there are apologists of the ``universal metal
dispersion'' calculating the LDA band structure and the FS and claiming the
rigid band behavior with Fermi level shift of the fixed band
dispersion~\cite{r32}. After the small hole pockets were discovered in the
Landau oscillations experiments~\cite{r10,r11}, the rigid band scenario becomes
evidently unconvincing. In place of conventional Fermi liquid state of the
normal metal, the pseudogap state appears in the phase diagram of cuprates
beside the long-range AFM phase. Though the origin of the pseudogap state is
still debated, the contribution of the fluctuating short-range AFM order is
clear~\cite{r5}. The short-range AFM order is essential not only in the UD
region. Even at the optimal doping the AFM correlation length $\xi _{AFM}
\approx 10$\AA ~\cite{r33}. At low temperatures $T \le 10K$, spin fluctuations
are slow with the typical time scale $10^{-9}$sec. and on the spatial scale of
$\xi_{AFM}$ (size of the AFM microdomain)~\cite{r34}. This time is large in
comparison to the fast electronic lifetime in ARPES ($\sim
10^{-13}$sec.)~\cite{r35} and to the cyclotron period $T \sim 2\pi
\omega_c^{-1} \sim 10^{-12}$ with $\omega_c$ being a cyclotron frequency in
quantum oscillations experiments~\cite{r10,r11}. Thus we safely consider that
the spin fluctuations are frozen at low $T$ and take into account only the
spatial dependence of the short-range AFM order. It means that the electronic
self-energy $\Sigma\left(k,\omega\right)$ will depends only on momentum,
$\Sigma\left(k,\omega\right) \to \Sigma(k)$.

We use this approach to study the concentration dependence of the electronic
structure and the FS. In Section~2 we present the electronic structure and the
change of the FS topology within $t-t'-t''-J^*$ model. The FS area and the
Luttinger theorem are also discussed. In Section~3 we give the qualitative
picture based on the interaction between hole and spin fluctuations. In
Section~4 we use the Lifshitz ideas~\cite{r36,r37} on the QPT to study the low
temperature thermodynamics. The electronic specific heat singularity near QPT
is compared to the experimental data~\cite{r38}.

\section{The Fermi surface of $La_{2-x}Sr_xCuO_4$ and its doping evolution}

Within the LDA+GTB approach we start from the {\it ab initio} LDA calculations
and construct the Wannier functions in the basis of oxygen $p$-orbitals and
copper $e_g$-orbitals. The multiband $p-d$ model~\cite{r39} parameters are
calculated {\it ab initio}. Then we apply the cluster perturbation
approach~\cite{r17,r40} and introduce the Hubbard $X$-operators constructed
within the full set of eigenstates of the unit cell (the CuO$_6$ cluster) that
is obtained by exact diagonalization of the multiband $p-d$ model Hamiltonian
of the cluster. By the GTB method we construct the low energy effective Hubbard
model with $U=E_{CT}$, where $E_{CT}$ is the change transfer gap~\cite{r41}. In
the Hubbard model, the $X_f^{0\sigma}$ operator describes the hole annihilation
at the site $f$ in the lower Hubbard band (LHB) of holes that corresponds to
the electron at the bottom of the conduction band. The hole annihilation in the
upper Hubbard band (UHB) is given by the $X_f^{\bar \sigma 2}$ operator and
corresponds to the electron at the top of the valence band. In the limit of
SEC, $U_{eff} \gg t$ ($t$ is the effective intersite hopping), we may exclude
either two-hole state $\left| 2 \right\rangle$ and obtain the effective
Hamiltonian for LHB or two-electron state $\left| 0 \right\rangle $ (hole
vacuum $d^{10}p^6$) and to get the effective Hamiltonian for UHB. Latter case
is interesting for the hole doped cuprates. We emphasize that the effective
$t-t'-t''-J^*$ model is derived from the microscopic approach and its
parameters are calculated {\it ab initio}. Here $J^*$ means that we take into
account the 3-site correlated hopping that is of the same order as the
superexchange interaction $J$.

The model Hamiltonian is given by
\begin{eqnarray}
H_{t-J^*} &=& H_{t - J} + H_{(3)}, \\
H_{t-J} &=& \sum\limits_{f,\sigma } {\left[ {(\varepsilon - \mu )X_f^{\sigma \sigma } +
(\varepsilon _2 - 2\mu )X_f^{22} } \right]} \nonumber\\
&+& \sum\limits_{f \ne g,\sigma } {t_{fg}^{11} } X_f^{2\bar \sigma } X_g^{\bar \sigma 2}
+ \sum\limits_{f \ne g} {J_{fg} \left( {\vec S_f \cdot \vec S_g - \frac{1}{4} n_f n_g } \right)} \nonumber\\
H_{(3)} &=& \sum\limits_{f \ne m \ne g,\sigma } {\frac{{t_{fm}^{01} t_{mg}^{01} }}
{{U_{eff} }}\left( {X_f^{\sigma 2} X_m^{\bar \sigma \sigma } X_g^{2\bar \sigma } -
X_f^{\bar \sigma 2} X_m^{\sigma \sigma } X_g^{2\bar \sigma } } \right)}. \nonumber
\end{eqnarray}
Here $J_{fg} = 2 \left(t_{fg}^{01}\right)^2/U_{eff}$, $t_{fg}^{01}$ is the
interband (LHB $\leftrightarrow$ UHB) hopping parameter between two sites $f$
and $g$, $\vec S_f $ is the spin operator, $\varepsilon$ and $\varepsilon_2$
are one- and two-hole local energies, and $\mu$ is the chemical potential. The
intraband hopping parameters $t_{fg}^{11}$ have been calculated up to 6-th
nearest neighbors. It appears that only 3 coordination spheres are important
and contribution to the hole dispersion of the more distant neighbors is
negligible. The calculated from {\it ab initio} model parameters for
La$_{2-x}$Sr$_x$CuO$_4$ are (in eV):
\begin{equation}
\begin{array}{l}
t = 0.932,\;t' = - 012,\;t'' = 0.152,\;\\
J = 0.298,\;J' = 0.003,\;J'' = 0.007.
\end{array}
\end{equation}

We introduce the hole Green function in the UHB (here $\bar\sigma \equiv
-\sigma$)
\begin{equation}
\label{green_func}
G_\sigma(\mathbf{k},E) = \left\langle \left\langle \left. X_\mathbf{k}^{\bar\sigma 2}
\right| X_\mathbf{k}^{2 \bar\sigma} \right\rangle \right\rangle_E.
\end{equation}
The analysis of the whole set of diagrams in the $X$-operators diagram
technique results in the following exact generalized Dyson equation~\cite{r42}
\begin{equation}
G_\sigma({\mathbf{k}},E) = \frac{{\rm P}_\sigma(\mathbf{k},E)}
{E - \epsilon_0 + \mu - {\rm P}_\sigma(\mathbf{k},E) t_\mathbf{k} - \Sigma_\sigma(\mathbf{k},E)}.
\end{equation}
Here $t_{\mathbf{k}}$ is the Fourier transform of the hopping, ${\rm
P}_\sigma(\mathbf{k},E)$ and $\Sigma_\sigma(\mathbf{k},E)$ are the strength and
the self-energy operators. In the simplest Hubbard I approximation
$\Sigma_\sigma = 0$, ${\rm P}_\sigma = F_{\bar\sigma 2} = \left\langle
X_f^{\bar\sigma \bar\sigma} \right\rangle + \left\langle X_f^{22}
\right\rangle$. The QP spectral weight is determined by the filling factor
$F_{\bar\sigma 2}$. In the diagram technique, $F_{\bar\sigma 2}$ corresponds to
the so-called ``terminal factors''~\cite{r43}.

To incorporate the effect of the short-range AFM order on the QP dynamics we go
beyond the Hubbard I approximation. The calculation scheme is given in
Ref.~\onlinecite{r44}. We use the Mori-type method to project the higher order
Green functions to the single particle function (\ref{green_func}). A similar
approach that took spin dynamics into the account was used in
Refs.~\onlinecite{r21,r45}. The hole concentration in La$_{2-x}$Sr$_x$CuO$_4$
(LSCO) per unit cell is $n_h = 1 + x$. The completeness condition for the local
Hilbert space in the $t-J$ model is
\begin{equation}
\sum\limits_\sigma X_f^{\sigma \sigma} + X_f^{22} = 1.
\end{equation}
Thus we easily obtain $\left\langle X_f^{\sigma \sigma} \right\rangle =
(1-x)/2$, $\left\langle X_f^{22} \right\rangle = x$, and $F_{\bar\sigma 2} =
(1+x)/2$. The Green function (\ref{green_func}) becomes
\begin{equation}
\label{eq:G}
G_\sigma(\mathbf{k},E) = \frac{(1+x)/2}
{E - \varepsilon _0 + \mu - \frac{1 + x}{2} t_{\mathbf{k}} - \frac{1 - x^2}{4}
\frac{\left( t_\mathbf{k}^{01} \right)^2}{U_{eff}} - \Sigma(\mathbf{k})},
\end{equation}
with the self energy given by
\begin{widetext}
\begin{eqnarray}
\Sigma(\mathbf{k}) &=& \frac{2}{1 + x}\frac{1}{N}
 \sum\limits_\mathbf{q} \left\{ \left[ t_\mathbf{q} - \frac{1 - x}{2}J_{\mathbf{k} - \mathbf{q}}
 - x \left(t_\mathbf{q}^{01} \right)^2/U_{eff} - (1 + x) t_\mathbf{k}^{01} t_\mathbf{q}^{01}/U_{eff} \right]
 K({\mathbf{q}}) \right. \\
&+& \left. \left[ t_{\mathbf{k} - \mathbf{q}} - \frac{1 - x}{2}\left(J_\mathbf{q}
 - \left(t_{\mathbf{k} - \mathbf{q}}^{01} \right)^2/U_{eff} \right)
 - (1 + x)t_\mathbf{k}^{01} t_{\mathbf{k} - \mathbf{q}}^{01}/U_{eff} \right]
 \cdot \frac{3}{2} C(\mathbf{q}) \right\}.
\end{eqnarray}
\end{widetext}
Here $K(\mathbf{q})$ and $C(\mathbf{q})$ stand for the Fourier transforms of
the static kinetic and spin correlation functions,
\begin{eqnarray}
\label{corr_func}
K(\mathbf{q}) &=& \sum\limits_{\mathbf{f} - \mathbf{g}} e^{-i(\mathbf{f} - \mathbf{g})\mathbf{q}}
 \left\langle X_\mathbf{f}^{2\bar \sigma} X_\mathbf{g}^{\bar\sigma 2} \right\rangle, \nonumber\\
C(\mathbf{q}) &=& \sum\limits_{\mathbf{f} - \mathbf{g}} e^{-i(\mathbf{f} - \mathbf{g})\mathbf{q}}
 \left\langle X_\mathbf{f}^{\sigma \bar\sigma} X_\mathbf{g}^{\bar\sigma \sigma} \right\rangle \nonumber\\
&=& 2\sum\limits_{\mathbf{f} - \mathbf{g}} e^{-i(\mathbf{f} - \mathbf{g})\mathbf{q}}
 \left\langle S_\mathbf{f}^z S_\mathbf{g}^z \right\rangle.
\end{eqnarray}

For the LHB which corresponds to the electron-doped cuprates, the similar Green
function has been obtained previously~\cite{r46}. We assume that the spin
system is an isotropic spin liquid with any averaged component of the spin
being zero and the equal correlation functions for any component of the spin,
$\left\langle S_f^\alpha S_g^\alpha \right\rangle$, $\alpha=x,\;y,\;z$. We
calculate this correlation function following Ref.~\onlinecite{r46} by the
method developed previously for the Heisenberg model~\cite{r47,r48}. The
resulting static magnetic susceptibility agrees with the other calculations for
the $t-J$ model~\cite{r49,r50}. As concerns the kinetic correlation function it
is expressed via the same Green function (\ref{green_func}).

\begin{figure*}
\includegraphics[clip=true,width=1.7\columnwidth]{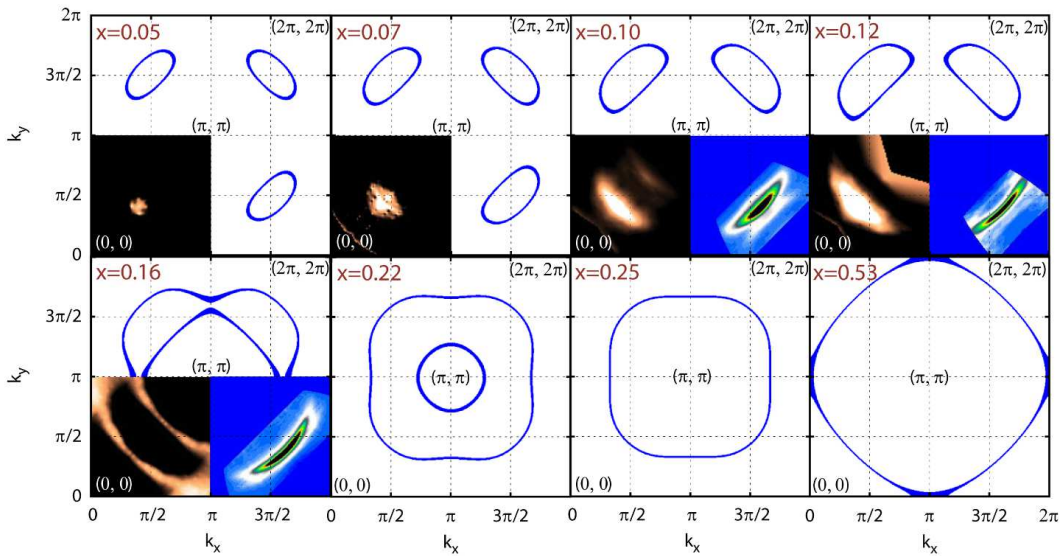}
\caption{\label{fig:1} Calculated Fermi surface for a single-layer cuprate for
different doping levels $x$. Fermi surface topology changes at $x_{c1} = 0.15$
and $x_{c2} = 0.24$. ARPES data from Ref.~\onlinecite{r51} and
Ref.~\onlinecite{r70} are shown in lower left and lower right corners of the
Brillouin zone, respectively.}
\end{figure*}

The self-consistent treatment of the electronic and spin systems results in the
evolution of the correlation functions (\ref{corr_func}), the chemical
potential, and the FS as function of doping (Fig.~\ref{fig:1}). At a small
doping we get 4 hole pockets close to $(\pi/2,\pi/2)$ point as was expected for
the AFM state. At the critical concentration $x_{c1} \approx 0.15$ the
connection of this pockets appears along the $(\pi,0)-(\pi,\pi)$ line and the
FS topology changes. At $x_{c1} < x < x_{c2} \approx 0.24$ we obtain two FS
centered around the $(\pi,\pi)$ point. The smaller one is the electronic FS; it
shrinks with doping and collapsed when $x \to x_{c2}$. The larger one is the
hole FS; with increasing $x$ it becomes more rounded. At $x > x_{c2}$ only a
large hole FS remains. Finally there is one more change of the topology at $x =
x_{c3}$ when the FS touches the $(\pi,0)$ point and becomes of electronic type
centered around the $(0,0)$ point.

Note that the values of critical concentrations are obtained with the finite
accuracy. First of all, model parameters are deduced by the complicated
procedure involving the projection of the LDA wave functions in Wannier
function basis and may vary with the change of this basis. Second, the equation
for the Green function (\ref{eq:G}) is approximate and taking into account
higher order corrections may change values of the critical concentrations
quantitatively. On the other hand, the qualitative picture should remain
unchanged since it is due to the general properties of the electron scattering
by the AFM fluctuations. Also, the qualitatively similar transformation of the
FS with doping has been obtained for the Hubbard model in the regime of SEC
(Fig.~15 in Ref.~\onlinecite{r45}), in the spin-density wave sate of the
Hubbard model~\cite{rSachdev}, for the spin-fermion model~\cite{r63}, and in
the {\it ab initio} multielectron quantum chemical approach~\cite{r68}.
Qualitative agreement of our results and results of calculations in different
approximations~\cite{r45,rSachdev,r63,r68} is basically due to the common
underlying idea: the change of the electron dispersion caused by the
interaction with the short-range AFM order. However, in our approach both
magnetic and electronic properties are treated self-consistently.

In Fig.~\ref{fig:1} we also show the ARPES data on Bi$_2$Sr$_{2 -
x}$La$_x$CuO$_{6 + y}$ (Bi2201) from Ref.~\onlinecite{r51} and the recent
data~\cite{r70} on LSCO for doping concentrations 0.10, 0.12, and 0.16. The
single crystals of Bi2201 have been studied experimentally with different hole
concentrations, $0.05 < p < 0.18$. This crystal has one CuO$_2$ layer in the
unit cell - that is why our calculations appropriate for LSCO can be used for
Bi2201 with the condition $x=p$. The question arise whether the model
parameters are the same or different for the two crystals? In the conventional
single electron tight-binding model used in Ref.~\onlinecite{r51} to fit the
ARPES data the hopping parameters depend on doping significantly. That is why
the authors of Ref.~\onlinecite{r51} claim that the ratio $t'/t$ is different
for Bi2201 and LSCO. However, as evident from figures 5a and 5b of
Ref.~\onlinecite{r51}, for the lowest doping the hopping values are close to
each other for both substances. The reason is that hopping parameters depend on
the interatomic distance that is almost the same in Bi2201 and LSCO. That is
why we use the same parameters for all doping concentrations. The doping
dependence of the band structure and its non-rigid behavior comes up as the
effect of SEC. One of the main players in this game is the filling factor
$F_{\bar\sigma 2}$.

Comparing our calculated FS with the experimental data in Fig.~\ref{fig:1} we
notice that for $x=$0.05, 0.07, 0.10, and 0.12 the experimental Fermi arc
position is close to the calculated inner part of the hole pocket (the part
which is near the $(0,0)$ point). The outer part appears as the small intensity
signal at $x=0.10$ and $x=0.12$ in ARPES. After the Lifshitz QPT at
$x_{c1}=0.15$ we see the two parts of the FS in agreement with ARPES
data~\cite{r51}. Usually the outer FS (the nearest to $(\pi,\pi)$ point) in
Bi-cuprates is ascribed as the superlattice reflection. It may be that the
superlattice signal simply mask the part of the FS that we obtain in the
calculation. Another most probable explanation is that the scattering by the
AFM fluctuations suppresses intensity of the spectral peaks corresponding to
the outer FS. We will discuss this scenario in the next Section.

At higher doping the ARPES results in the large hole pocket centered around the
$(\pi,\pi)$ point~\cite{r52}, \textit{e.g.} in Tl$_2$Ba$_2$CuO$_{6 + y}$ with
$p=0.26$. Our calculations result in such topology for $x > x_{c2}$. According
to Ref.~\onlinecite{r66}, there is an electron pocket for LSCO at $x=0.30$.

\begin{figure}
\includegraphics[clip=true,angle=270,width=0.97\columnwidth]{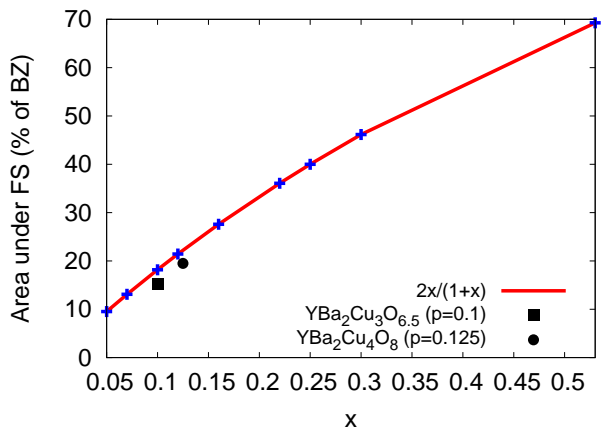}
\caption{\label{fig:2} The doping dependence of the FS area (in \% of the
Brillouin zone area) calculated directly (+), from the generalized Luttinger
theorem (solid line). The experimental values from the quantum oscillations
data~\cite{r10,r11} are also shown.}
\end{figure}
Now we would like to discuss the FS area and the Luttinger theorem. In
Fig.~\ref{fig:2} we give the FS area as a function of doping. Note that the
standard formulation of the Luttinger theorem does not work for Hubbard
fermions. For free electrons each quantum state in the $k$-space contains 2
electrons with opposite spins. The spectral weight of the Hubbard fermion is
determined by the strength operator, ${\rm P}_\sigma = F_{\bar\sigma 2}$, and
each quantum state contains $2 F_{\bar\sigma 2} = 1 + x$ electrons. A
generalized Luttinger theorem for SEC system~\cite{r67} takes into account the
spectral weight of each $\left| k \right\rangle$ state. For LSCO the hole
concentration $n_h = 1 + x$, so the electron concentration $n_e = 1 - x$. Using
the dispersion law (see Fig.~\ref{fig:3}b below) we calculate the number of
occupied electronic states $N_k^e$ below the Fermi level. The electronic
concentration $n_e = 2 F_{\bar\sigma 2} N_k^e = 1 - x$. It gives us $N_k^e = (1
- x)/(1 + x)$. Then the number of free (occupied by holes) $k$-states is $N_k^h
= 1 - N_k^e = 2x/(1 + x)$, and the FS area in Fig.~\ref{fig:2} is determined
by this number. The FS area obtained by direct calculation of the occupied
$k$-state under the Fermi level is shown by crosses. Two available FS areas
from the quantum oscillations data~\cite{r10,r11} are also marked in the
Fig.~\ref{fig:2}. It is evident and very important that the Luttinger theorem
is not applicable in the standard formulation. On the other hand, its
generalization for the case of correlated Hubbard fermions describes the
experimental data very well.

\section{Qualitative analysis of the electron dispersion and ARPES in a system with the short-range AFM background}

It was shown earlier~\cite{r12,r13,r14,r15} that AFM fluctuations transform the
closed hole pocket into the arc. We will extent the same arguments to the
doping region where AFM fluctuations are strong. The electron Green function on
the square lattice with electron scattering by the Gaussian fluctuations that
imitate the short-range AFM order with $Q = (\pi,\pi)$ is equal to~\cite{r15}
\begin{equation}
\label{eq_9}
G_D(\mathbf{k},E) = \frac{E - \varepsilon(\mathbf{k} + \mathbf{Q}) + i v k}
{\left(E - \varepsilon(\mathbf{k}) \right)\left(E - \varepsilon(\mathbf{k} + \mathbf{Q}) + i v k \right) - |D|^2}.
\end{equation}
Here $|D|$ stands for the amplitude of the fluctuating AFM order,
$\varepsilon(\mathbf{k})$ is the electron energy in the paramagnetic phase,
\begin{equation}
v = \left| v_x(\mathbf{k} + \mathbf{Q})\right| + \left| v_y (\mathbf{k} + \mathbf{Q}) \right|,\;
v_{x,y}(\mathbf{k}) = \frac{\partial \varepsilon(\mathbf{k})}{\partial k_{x,y}}.
\end{equation}

In the absence of the damping the Green function (\ref{eq_9}) describes the
electron in the spin-density wave state with the long-range order where Umklapp
shadow band is given by $\varepsilon(\mathbf{k} + \mathbf{Q})$. On then other
hand, for the AFM spin-liquid there is a dynamical transition
$\varepsilon(\mathbf{k}) \to \varepsilon(\mathbf{k} + \mathbf{Q})$ with finite
lifetime $1/\tau \sim v\mathbf{k}$.

\begin{figure}
\begin{minipage}[h]{1\columnwidth}
\center{\includegraphics[clip=true,angle=270,width=0.88\columnwidth]{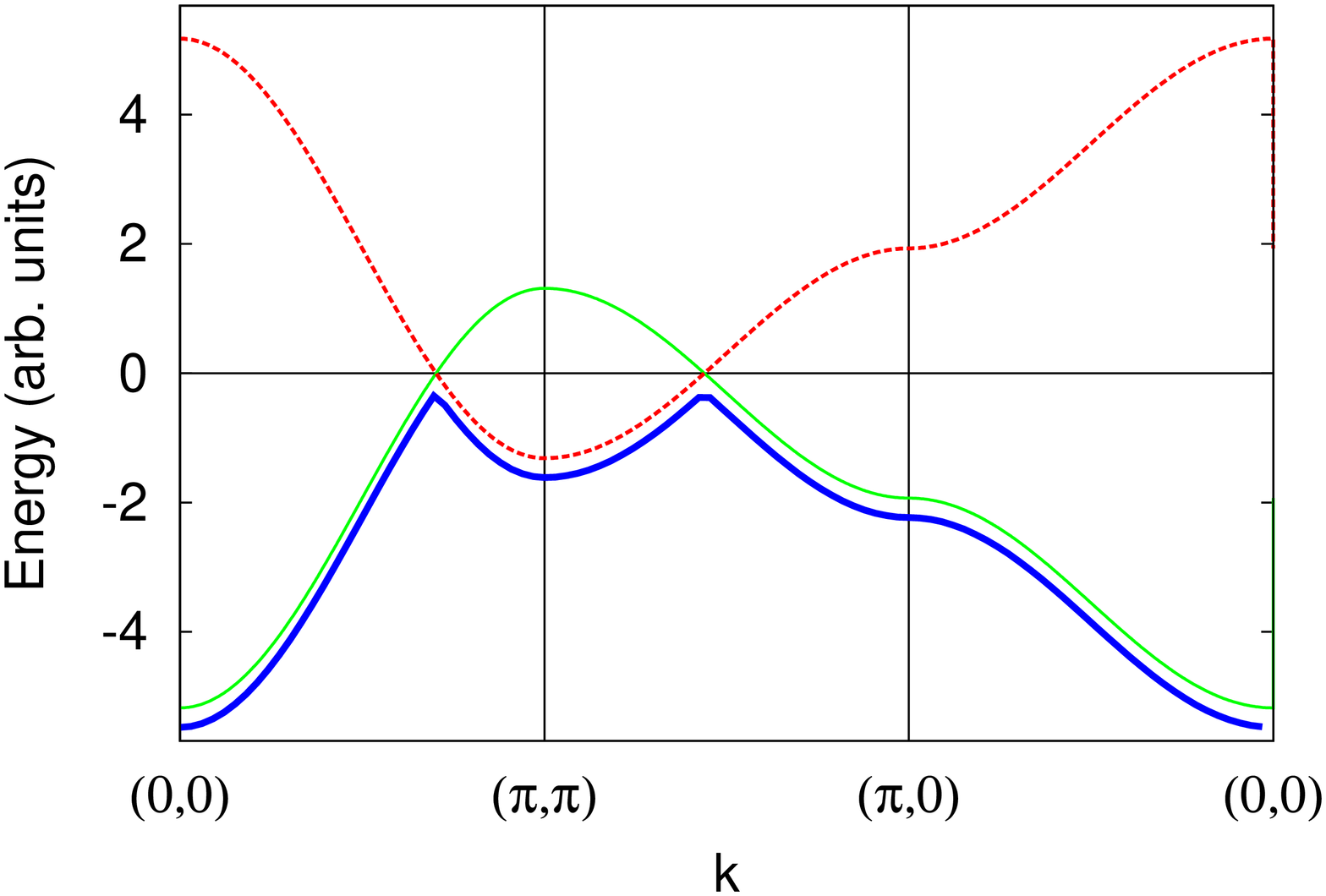}} (a) \\
\end{minipage}
\begin{minipage}[h]{1\columnwidth}
\center{\includegraphics[clip=true,angle=270,width=0.9\columnwidth]{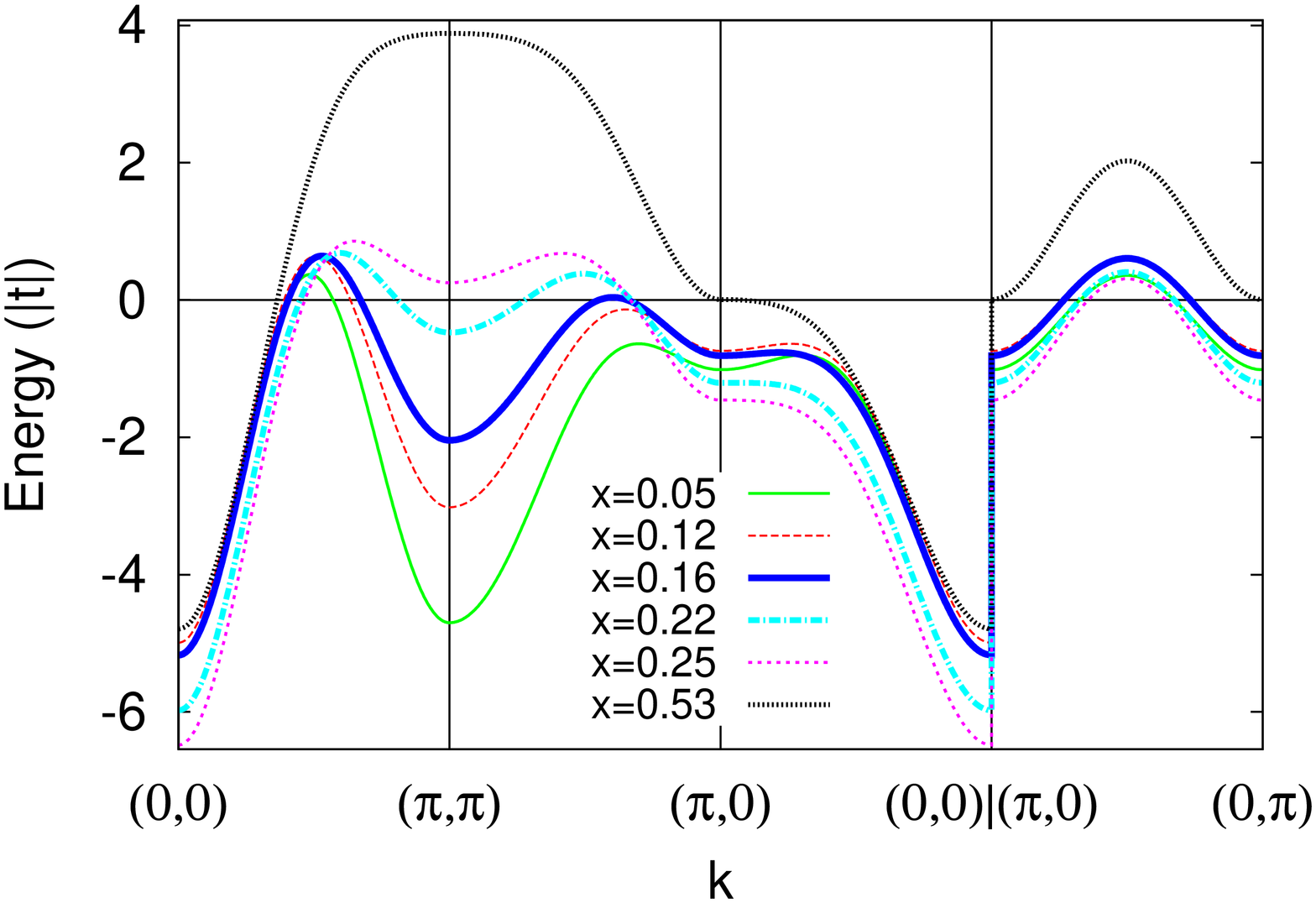}} (b) \\
\end{minipage}
\begin{minipage}[h]{1\columnwidth}
\center{\includegraphics[clip=true,width=0.7\columnwidth]{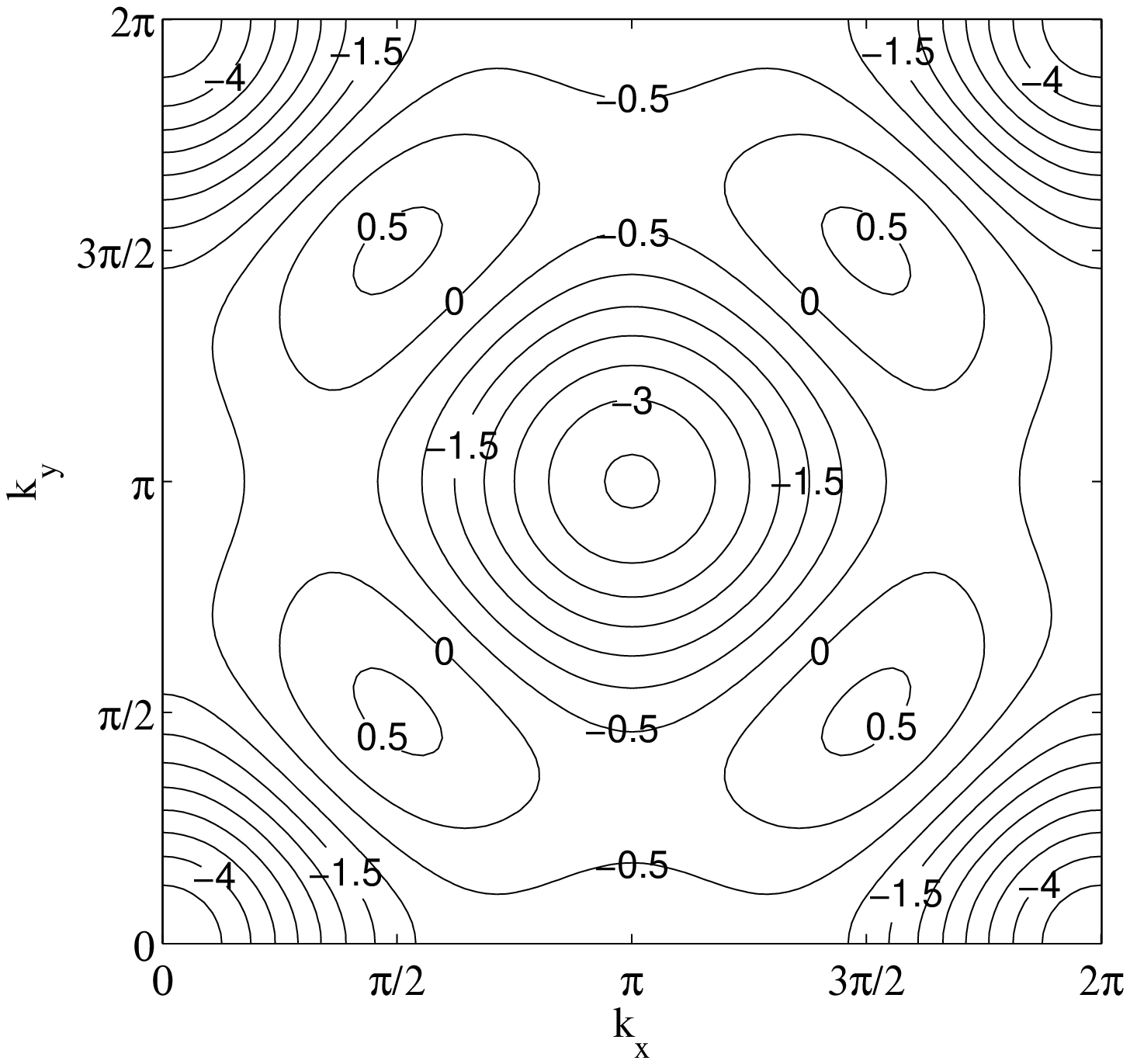}} (c) \\
\end{minipage}
\caption{\label{fig:3} A qualitative scheme of the band structure of the electron on the
fluctuating AFM background (a), our calculations (b), and the constant-energy cuts
for $x = 0.10$ (c). The zero energy in (b) and (c) corresponds to the Fermi level.
The constant energy contours in (c) are labeled by
the values of the corresponding energies (in units of $t$).}
\end{figure}
The paramagnetic dispersion is shown in Fig.~\ref{fig:3}a by a thin green curve
and a shadow band by a dotted curve to stress that it has the finite lifetime
as follows from equation (\ref{eq_9}). The resulting QP dispersion in the
short-range AFM state is given by a thick blue curve. With increasing doping
the Fermi level moves down from its initial value ``0'' in the
Fig.~\ref{fig:3}a. The first intersection occurs along the $(0,0) - (\pi,\pi)$
direction and results in 4 small hole pockets. The inner part of the FS is
formed mainly by the non-damped electrons from the $\varepsilon(\mathbf{k})$
band while the outer part is formed mainly by the damped shadow band. That is
why the outer part has a very small spectral weight and was not seen in ARPES
data until recent discovery by the laser ARPES with the ultra-high energy
resolution~\cite{r70} (see Fig.~\ref{fig:1}). This qualitative analysis
reproduce the calculations~\cite{r12,r13,r14,r15,r45}.

Now we proceed to higher doping concentrations. For $x=0.16$ where AFM
correlation length $\xi_{AFM} \approx 10$\AA we have two large FS centered
around $(\pi, \pi)$. Those can be deduced from Fig.~\ref{fig:3}a by further
decrease of the Fermi energy, $\mu$. The critical point $x_{c1}$ appears when
$\mu$ touches the second peak along the $(\pi, \pi) - (\pi, 0)$ direction. It
is clear from Fig.~\ref{fig:3}a that the inner FS will be of the electronic
type and is formed by the damped shadow band. Thus the corresponding spectral
peaks are very small. The outer FS is of the hole type and is formed by the
non-damped states. That is why its intensity is much larger than that of the
inner part (see Ref.~\onlinecite{r51} and Fig.~\ref{fig:1} for $x=0.16$). With
further decrease of $\mu$ it will cross the bottom of the band at $x = x_{c2}$
and that corresponds to the collapse of the electronic FS. Finally at $x
> x_{c2}$ the crossing of $\mu$ with the saddle point at $(\pi, 0)$ results in
the transformation of the FS from the hole to the electron type at $x =
x_{c3}$. Latter takes place in a strongly OD regime; this effect can be
obtained in any conventional single electron approach and has been discussed
before~\cite{r69}. For comparison, we present our calculated band structure for
various doping concentration in Fig.~\ref{fig:3}b and the constant energy cut
in Fig.~\ref{fig:3}c. It is clear that the rigid band approach of
Fig.~\ref{fig:3}a may give the correct sequence of the FS reconstruction but
quantitatively it is wrong.

\section{Low temperature thermodynamics near the Lifshitz transition}

According to Lifshitz results~\cite{r36,r37}, both FS transformations at
$x_{c1}$ and $x_{c2}$ are 2.5 order electronic phase transitions (nowadays the
term QPT is used). Appearance of a new FS sheet at $\varepsilon =
\varepsilon_c$ gives the additional density of state $\delta g(\varepsilon) =
\alpha (\varepsilon - \varepsilon_c )^{1/2}$, with $\alpha \sim 1$ in a 3D
system. In spite of a strong anisotropy in cuprates they are 3D crystals. Weak
interlayer hopping results in a FS modulation along the $k_z$ axis that has
been measured by ARPES~\cite{r32}. That is why we can use results of
Refs.~\onlinecite{r36,r37} with minimal modification due to the QP spectral
weight in the strongly correlated system $F_{\bar\sigma 2} = (1 + x)/2$.

Near the critical point the thermodynamical potential gains additional
contribution:
\begin{equation}
\Omega(\mu, T) = \Omega_0 (\mu, T) + \delta\Omega.
\end{equation}
This singular contribution is induced by a new FS sheet at $\varepsilon >
\varepsilon_c $ and is equal to
\begin{equation}
\delta\Omega = - \int\limits_0^\infty {\delta N(\varepsilon) f_F(\varepsilon) d\varepsilon},
\end{equation}
where $f_F(\varepsilon)$ is the Fermi function. The number of states is given
by
\begin{equation}
\delta N(\varepsilon) = \left\{ \begin{array}{l}
                           0, \varepsilon < \varepsilon_c \\
                           \frac{2}{3} \alpha \frac{1 + x}{2} (\varepsilon - \varepsilon_c )^{3/2}, \varepsilon > \varepsilon_c
                         \end{array} \right.
\end{equation}
At low temperature, $T \ll z$, $z = \mu - \varepsilon_c$, and close to the QPT
at $z = 0$ we get
\begin{equation}
\delta\Omega = \left\{ \begin{array}{l}
   - \frac{\sqrt\pi}{4} (1 + x) \alpha T^{5/2} e^{-|z|/T}, z < 0 \\
   - \frac{2}{15} (1 + x) \alpha |z|^{5/2} - \frac{\pi^2}{12} (1 + x) T^2 |z|^{1/2}, z > 0
 \end{array} \right.
\end{equation}
It is a $z^{5/2}$ singularity that tells about 2.5 phase transition. In our
case $z$ depends on doping so $z(x) = 0$ at $x = x_{c1}$ and $x = x_{c2}$.

The singular contribution to the Sommerfeld parameter $\gamma = C_e/T$ where
$C_e$ is the electronic specific heat, has the following form
\begin{eqnarray}
\delta\gamma &=& - \frac{\partial^2 \delta F}{\partial T^2} \\
&=& \left\{ {\begin{array}{l}
   \frac{\sqrt\pi}{4} (1 + x) \alpha \frac{|z|^2}{T^2}
    \left( 1 + 3\frac{T}{|z|} + \frac{15}{4}\frac{T^2}{|z|^2} \right) e^{-|z|/T}, z < 0 \\
   \frac{\pi^2}{6} (1 + x) \alpha z^{1/2}, z > 0
 \end{array} } \right. \nonumber
\end{eqnarray}
\begin{figure}
\includegraphics[angle=270,width=0.97\columnwidth]{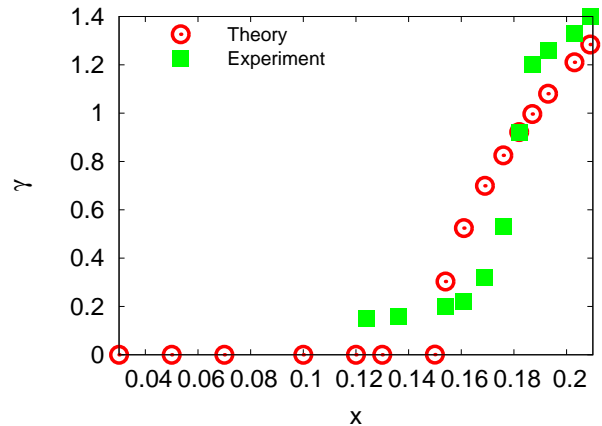}
\caption{\label{fig:4} The Sommerfild parameter near the Lifshitz QPT. Experimental data for
$\gamma=C_e/T$ at $T = 10 K$ were taken from Ref.~\onlinecite{r38}.}
\end{figure}
We have deduced $z(x)$ dependence near each critical point from our band
structure calculations. Obtained $\delta\gamma$ at $T = 10 K$ near $x_{c1}$ is
shown in the Fig.~\ref{fig:4}. We also plot the experimental data~\cite{r38}
for LSCO, where $C_e$ has been obtained by extrapolation of the high
temperature data for $T > T_c$ to the low-$T$ region. The experimental points
in Fig.~\ref{fig:4} correspond to the total $\gamma$,
\begin{equation}
\gamma(x) = \gamma_0(x) + \delta\gamma,
\end{equation}
where $\gamma_0$ is a smooth function at $x \approx x_{c1}$.

Since the electron FS pocket disappears for $x > x_{c2}$, for $x < x_{c2}$ our
theory produce a singular behavior of $\gamma(x)$ corresponding to the case of
$z > 0$. Measurements of the electronic specific heat~\cite{r53} in
NdBa$_2$Cu$_3$O$_{6 + y}$ revealed two weak maxima of $\gamma(x)$ at $p = 0.16$
and $p = 0.23$ that are close to our $x_{c1}$ and $x_{c2}$. To stay away from
the superconductivity, measurements of Ref.~\onlinecite{r53} were carried out
at $T = 200 K$ that explains why singularities appear as weak maxima.

\section{Conclusion}

Previously, transformations of the FS has been discussed within a variational
approach to the $t-J$ model~\cite{r27}. The small hole pocket near the
$(\pi/2,\pi/2)$ point has been obtained in the UD AFM. At large doping, the
electronic FS around $(0, 0)$ point also has been obtained. Nevertheless, the
FS for intermediate $x$ in Ref.~\onlinecite{r27} does not correspond to our FS
and to the experimental data.

Recently there were a lot of discussions on the change of the carrier sign upon
doping. At large $x$, the FS becomes of the electronic type: in LSCO it happens
at $x > 0.30$~\cite{r54}. As was mentioned above, it is rather a trivial fact.
More unusual are the experimental data on the change of the Hall coefficient
($R_H$) sign in the UD systems. This effect was observed (under a strong
magnetic field of $50 \div 60$T that suppress the superconductivity) in
YBa$_2$Cu$_3$O$_y$ with $p =$0.10, 0.12 and 0.14~\cite{r55}, and in LSCO with
$p = 0.11$~\cite{r56}. All these crystals belong to the region $x < x_{c1}$ and
according to our theory should have the small hole FS pockets. We believe that
the arguments of Ref.~\onlinecite{r57} may explain the negative total Hall
coefficient due to opposite partial contributions to $R_H$ of the FS with
opposite curvatures in the two-dimensional metal.

Low temperature transport measurements on La$_{1.6 - x}$Nd$_{0.4}$Sr$_x$CuO$_4$
in a strong magnetic field up to $35$T reveal the change of the FS topology at
$p^* \approx 0.23$~\cite{r58}. This critical point is very close to our $x_{c2}
= 0.24$. Also, our theory agrees with the data of Ref.~\onlinecite{r58} in the
sense that at $p = 0.24$ the $R_H$ indicates the large cylindrical FS with $1 +
p$ holes. At $p = 0.20$ that corresponds to $x < x_{c2}$, the $R_H(T)$ increase
at low temperature leads to the conclusion that the FS reconstruction and
pseudogap formation happen at $p < p^*$~\cite{r58}. The critical concentration
$x_{c2}$ agrees with the concentration $p_c = 0.23$ where the van Hove
singularity in Bi2201 have been found in ARPES~\cite{r59,r60}.

There is a wide discussion in the literature on the quantum critical point
$P_{crit}$ where the pseudogap characteristic temperature $T^*(P) \to 0$.
According to Ref.~\onlinecite{r61}, $P_{crit} = 0.19$ and according to
Ref.~\onlinecite{r62}, $P_{crit} = 0.27$. All these values are obtained by
extrapolation from finite-$T$ regime. On the contrary, the two critical points
$x_{c1}$ and $x_{c2}$ obtained here are the properties of the ground state and
results from the Lifshitz QPT. It is well maybe that our $x_{c2}$ is somehow
related to the $P_{crit}$, at least $p^* = 0.24$ is related in
Ref.~\onlinecite{r58} to the pseudogap formation at $p < p^*$.

\begin {acknowledgments}
We thanks A. Kordyuk for the discussion of results and T.M. Ovchinnikova for
technical assistance. This work was supported by project 5.7 of the Presidium
of RAS programm ``Quantum physics of the condenced matter'', RFFI grant
09-02-00127, and integration project N 40 of SB RAS.
\end {acknowledgments}


\begin {thebibliography}{28}
\bibitem{r1} E. Dagotto, Rev. Mod. Phys. {\bf 66}, 763 (1994)
\bibitem{r2} E.G. Maksimov, Phys. Usp. {\bf 43} 965 (2000)
\bibitem{r3} M. Imada, A. Fujimori, Y. Tokura, Rev. Mod. Phys. {\bf 70}, 1039 (1998)
\bibitem{r4} S.G. Ovchinnikov, Phys. Usp. {\bf 40} 993 (1997)
\bibitem{r5} M.V. Sadovskii, Phys. Usp. {\bf 44} 515 (2001)
\bibitem{r6} V.F. Elesin, V.V. Kapaev, Yu.V. Kopaev, Phys. Usp. {\bf 47} 949 (2004)
\bibitem{r7} Yu.A. Izyumov, E.Z. Kurmaev, Phys. Usp. {\bf 51} 23 (2008)
\bibitem{r8} P.A. Lee, Rep. Prog. Phys. {\bf 71}, 012501 (2008)
\bibitem{r9} A. Damascelli, Z. Hussein, Z.X. Shen, Rev. Mod. Phys. {\bf 75}, 473 (2003)
\bibitem{r10} N. Doiron-Leyrand et al, Nature {\bf 447}, 565 (2007)
\bibitem{r11} E.A. Yelland, J.Singleton, C.H. Mielke, N. Narrison, F.F. Balakirev, B. Dabrowski, J.R. Cooper, Phys. Rev. Lett. {\bf 100}, 047003 (2008)
\bibitem{r12} E.Z. Kuchinskii, I.A. Nekrasov, M.V. Sadovskii, JETP Lett. {\bf 82}, 198 (2005)
\bibitem{r13} E.Z. Kuchinskii, M.V. Sadovskii, JETP {\bf 103}, 415 (2006)
\bibitem{r14} N. Harrison, R.D. McDonald, J. Singleton, Phys. Rev. Lett. {\bf 99}, 206406 (2007)
\bibitem{r15} E.Z. Kuchinskii, M.V. Sadovskii, JETP Lett. {\bf 88}, 192 (2008)
\bibitem{r70} J. Meng, G. Liu, W. Zhang et al., arXiv:0906.2682 (2009)
\bibitem{r16} M.Yu. Kagan, K.I. Kugel, Phys. Usp. {\bf 44} 553 (2001)
\bibitem{r17} S.G. Ovchinnikov, I.S. Sandalov, Physica C {\bf 161}, 607 (1989)
\bibitem{r18} S.V. Lovtsov, V.Yu. Yushankhai, Physica C {\bf 179}, 159 (1991)
\bibitem{r19} J.H. Jefferson, H. Eskes, L.F. Feiner, Phys. Rev. B {\bf 45}, 7959 (1992)
\bibitem{r20} V.I. Belinicher, A.L. Chernyshev, V.A. Shubin, Phys. Rev. B {\bf 53}, 335 (1996)
\bibitem{r21} N.M. Plakida, V.S. Oudovenko, Phys. Rev. B {\bf 59}, 11949 (1999)
\bibitem{r22} M.M. Korshunov, V.A. Gavrichkov, S.G. Ovchinnikov, I.A. Nekrasov, Z.V. Pchelkina, V.I. Anisimov, Phys. Rev. B {\bf 72}, 165104 (2005)
\bibitem{r23} W. Stephan, P. Horsch, Phys. Rev. Lett. {\bf 66}, 2258 (1991)
\bibitem{r24} R. Preuss, W. Hanke, W. von der Linden, Phys. Rev. Lett. {\bf 75}, 1344 (1995)
\bibitem{r25} V.F. Elesin, V.A. Koshurnikov, J. Exp. Theor. Phys. {\bf 79}, 961. (1994)
\bibitem{r26} B.I. Shraiman, E.D. Siggia, Phys. Rev. Lett. {\bf 61}, 467 (1988)
\bibitem{r27} S.A. Trugman, Phys. Rev. Lett. {\bf 65}, 500 (1990)
\bibitem{r28} A.F. Barabanov, Superconductivity: Physics, Chemistry and
    Technology {\bf 3}, 8 (1990) [in Russian]
\bibitem{r29} A.F. Barabanov, R.O. Kuzian, L.A. Maksimov, J. Phys.: Condens. Matter. {\bf 39}, 129 (1991)
\bibitem{r30} A.P. Kampf, Phys. Rev. {\bf 249}, 219 (1994)
\bibitem{r31} G. Dorf, A. Muramatsu, W. Hanke, Phys. Rev. B {\bf 41}, 9264 (1990)
\bibitem{r32} S. Sahrakorpi, R.S. Markiewicz, H. Lin et al, Phys. Rev. B {\bf 78}, 104513 (2008)
\bibitem{r33} T.R. Thurston, R.J. Birgeneau, M.A. Kastner et al, Phys. Rev. B {\bf 40}, 4585 (1989)
\bibitem{r34} S.M. Haden et al, Phys. Rev. Lett. {\bf 66}, 821 (1991)
\bibitem{r35} D. Mihailovic, V.V. Kabanov, Superconductivity in Complex Systems. Series: Structure and Bonding, Vol. {\bf 114}, edited by K. A. Muller and A. Bussmann-Holder (Springer Verlag, Berlin, 2005), p. 331
\bibitem{r36} I.M. Lifshitz, Sov. Phys. JETP {\bf 11} 1130 (1960)
\bibitem{r37} I.M. Lifshitz, M.Ya. Asbel and M.I. Kaganov Electron Theory of Metals, Consultant Bureau, New York (1973)
\bibitem{r38} J.W. Loram, J. Luo, J.R. Cooper, W.Y. Liang, J.L. Tallon, Phys. Chem. Solids {\bf 62}, 59 (2001)
\bibitem{r39} Ya.B. Gaididei, V.M. Loktev. Phys. Stat. Sol. B {\bf 147}, 307 (1988)
\bibitem{r40} V.A. Gavrichkov, S.G. Ovchinnikov, À.À. Borisov, E.G. Goryachev, JETP {\bf 91} 369 (2000)
\bibitem{r41} J. Zaanen, G.A. Sawatzky, J.W. Allen, Phys. Rev. Lett. {\bf 55}, 418 (1985)
\bibitem{r42} S.G. Ovchinnikov, V.V. Val'kov, Hubbard operators in the Theory of Strongly correlated electrons, Imperial College Press, London-Singapure, 2004.
\bibitem{r43} R.O. Zaitsev, Sov. Phys. JETP {\bf 41}, 100 (1975)
\bibitem{r44} M.M. Korshunov, S.G. Ovchinnikov, Eur. Phys. J. B {\bf 57}, 271 (2007)
\bibitem{r45} N.M. Plakida, V.S. Oudovenko, JETP {\bf 104} 230 (2007)
\bibitem{r46} V.V. Val'kov, D.M. Dzebisashvili, JETP {\bf 100} 608 (2005)
\bibitem{r47} H. Shimahara, S. Takada, J. Phys. Soc. Jpn {\bf 60}, 2394 (1991); {\bf 61}, 989 (1992)
\bibitem{r48} A.F. Barabanov, V.M. Berezovskii, JETP {\bf 79} 627 (1994)
\bibitem{r49} A. Sherman. M. Schreiber, Phys. Rev. B {\bf 65}, 134520 (2002)
\bibitem{r50} A.A. Vladimirov, D. Ihle, N.M. Plakida, Theor. Math. Phys. {\bf 145}, 1576 (2005)
\bibitem{r51} M. Hashimoto, T. Yoshida, H. Yagi et al, Phys. Rev. B {\bf 77}, 094516 (2008)
\bibitem{r52} M. Plate, J.D.F. Mottershead, I.S. Elfimov et al, Phys. Rev. Lett. {\bf 95}, 077001 (2005)
\bibitem{r53} U. Tutsch, P. Schweiss, H. W?he, B. Obst, Th. Wolf, Eur. Phys. J. B {\bf 41}, 471 (2004)
\bibitem{r54} I. Tsukada, S. Ono, Phys. Rev. B {\bf 74}, 134508 (2006)
\bibitem{r55} D.Le Boeuf, N. Doiron-Leyraud, J. Levallois et al, Nature {\bf 450}, 533 (2003)
\bibitem{r56} T. Adachi, T. Noji, Y. Koike, Phys. Rev. B {\bf 64}, 144524 (2001)
\bibitem{r57} N.P. Ong, Phys. Rev. B {\bf 43}, 193 (1991)
\bibitem{r58} R. Daou, N. Doiron-Leyrand, D. Le Boeuf et al, Nature Phys. 5, {\bf 31} (2009)
\bibitem{r59} A. Kaminski, S. Rosenkranz, N.M. Fretweel et al, Phys. Rev. B {\bf 73}, 174511 (2006)
\bibitem{r60} A.A. Kordyuk, S.V. Borisenko, M. Khupfer, J. Fink, Phys. Rev. B {\bf 67}, 064504 (2003)
\bibitem{r61} J.G. Storey, J.L. Tallon, G.V.M. Williams, Phys. Rev. B {\bf 78}, 140506(R) (2008)
\bibitem{r62} S. Hufner, M.A. Hossain, A. Damascelly, G.A. Sawatzky, Rep. Prog. Phys. {\bf 71}, 062501 (2008)
\bibitem{rSachdev} S. Sachdev, A.V. Chubukov, A. Sokol, Phys. Rev. B {\bf 51}, 14874 (1995)
\bibitem{r63} A.F. Barabanov, A.A. Kovalev, O.V. Urusaev, A.M. Belemuk, R.
    Hain, JETP {\bf 92} 677 (2001)
\bibitem{r68} L. Hozoi, M.S. Laad, P. Fulde, Phys.Rev.B {\bf 78}, 165107 (2008)
\bibitem{r67} M.M. Korshunov, S.G. Ovchinnikov, Phys. Sol. St. {\bf 45}, 1415 (2003)
\bibitem{r66} A. Ino, C. Kim, M. Nakamura \textit{et al.}, Phys.Rev. B {\bf
    65}, 094504 (2002)
\bibitem{r69} F. Onufrieva, P. Pfeuty, Phys.Rev.B {\bf 61}, 799 (2000)
\end {thebibliography}
\end {document}